\documentstyle[11pt,paspconf,psfig]{article}

\begin{document}

\title{A radio survey of merging clusters in the Shapley Concentration}

\author{T.Venturi, R. Morganti}
\affil{Istituto di Radioastronomia, CNR, via Gobetti 101, I--40129 Bologna, 
Italy}

\author{S. Bardelli}
\affil{Osservatorio Astronomico di Trieste, via Tiepolo 11, I--34131 Trieste, 
Italy}
\author{D. Dallacasa}
\affil{Dipartimento di Astronomia, via Ranzani 1, I--40127 Bologna,
Italy}

\author{R.W. Hunstead}
\affil{School of Physics, University of Sydney, NSW 2006, Australia}

\begin{abstract}

We present here the first results of a 22 cm survey of the Shapley 
Concentration core. The observations were carried out with the
Australia Telescope Compact Array. Our radio observations completely
and uniformely cover the A3558 complex, allowing a thorough
multifrequency study, by comparison of our results with the available
optical spectroscopic and X--ray data of the whole chain.
We will present here some statistical results of our survey and compare
them with the information on the dynamics of the chain and on the
properties of the intracluster gas. Attention will also be devoted to
the extended radio galaxies found in our survey.

\end{abstract}



\section{Introduction}

Rich superclusters are the ideal environment for the detection of cluster 
mergings, because the peculiar velocities induced by the enhanced local density
of the large scale structure favour the cluster-cluster and cluster-group 
collisions, in the same way as the caustics seen in the simulations. 
The most remarkable examples of cluster merging seen at an intermediate stage
are found in the central region of the Shapley Concentration, the richest
supercluster of clusters found within 300 h$^{-1}$ Mpc (Zucca et al. 1993;
hereafter h=H$_o$/100). On the basis
of the two-dimensional distribution of galaxies of the COSMOS/UKSTJ catalogue,
it is possible to find several complexes of interacting
clusters, which suggest that the entire
region is dynamically active. Therefore, this supercluster represents a unique 
laboratory where it is possible to follow cluster mergings and to test related 
astrophysical consequences. It is believed that such dynamical
events may modify the properties of the emission of cluster galaxies
favouring 
the formation of radio halos, relicts and wide angle tail radio sources and 
the presence of post-starburst (E+A) galaxies.

\section{The A3558 cluster complex}

The A3558 cluster complex is a remarkable chain formed by the ACO 
clusters A3558, A3562 and A3556, located at $\sim 14500$ km/s and
spanning $\sim 7$ h$^{-1}$ Mpc perpendicular to the line of sight 
(see Fig. 1) 
This structure is approximately at the geometrical centre of the Shapley 
Concentration and can be considered the core of this supercluster.

%
\begin{figure}
\centerline{\vbox{
\psfig{figure=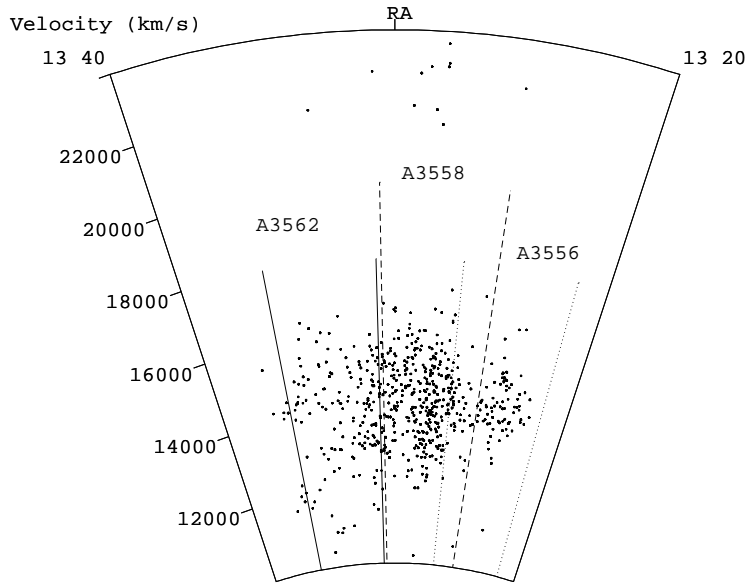,width=0.8\hsize}
}}
\caption[]{
Wedge diagram of the sample of galaxies in the velocity range 
$10000 - 24000$ km/s. The coordinate range is 
$13^h 22^m 06^s < \alpha (2000)< 13^h 37^m 15^s$ and 
$-32^o 22' 40''< \delta(2000) < -30^o 59' 30''$. The three pairs of straight 
lines (solid, dashed and dotted) show the projection in right ascension of
1 Abell radius for the three clusters A3562, A3558 and A3556, respectively.}
\end{figure}
%
\begin{figure}
\centerline{\vbox{
\psfig{figure=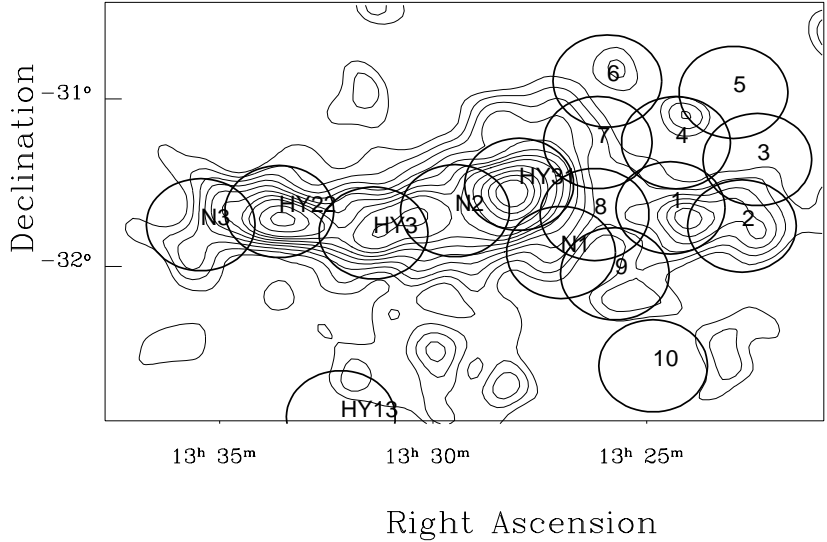,width=0.9\hsize}
}}
\caption[]{
Contours of the galaxy density in the A3558 complex. The centres of the 
superimposed circles correspond to the pointing centres of the
observations, and the circle radius is the primary beam of ATCA at
22 cm. Pointings 1 to 10 are presented in Venturi et al. 1997. The pointings
labelled HY31, HY3, HY13 and HY22 correspond to the archive data.}
\end{figure}
%

By the use of multifibre spectroscopy, we obtained 714 redshifts of galaxies 
in this region, confirming that the complex is a single connected structure,
elongated perpendicularly to the line of sight. In particular, the number of 
measured redshifts of galaxies belonging to A3558 is $307$, and thus this
is one of the best sampled galaxy clusters in the literature.
Moreover, two smaller 
groups, dubbed SC 1327-312 and SC 1329-313, were found both in the optical 
(Bardelli et al 1994) and X-ray band (Bardelli et al 1996).

After a substructure analysis, Bardelli et al. (1998) found that a large
number of subgroups reside in this
complex, meaning that this structure is dynamically active.
It is possible to consider two main hypothesis
for the origin of the A3558 cluster complex: one is that we are observing a 
cluster-cluster collision, just after the first core-core encounter, while 
the other considers repeated, incoherent group-group and group-cluster mergings
focused  at the geometrical centre of a large
scale density excess (the Shapley Concentration).
The second hypothesis seems to be favoured by the presence of an excess of blue
galaxies between A3558 and A3562, i.e. in the position where the shock is 
expected.

\section{The radio survey}

In order to test the effects of merging and group collision on the
radio properties of clusters of galaxies, and on the radio properties
of the galaxies within merging clusters, we started an extensive radio
survey of the A3556-A3558-A3562 chain (hereinafter the A3558 complex)
in the Shapley Concentration core. Our survey is being carried out at
22/13 cm with the Australia Telescope Compact Array (ATCA).
The main aims
of our study can be summarised as follows:

\noindent
{\it (a)} derive the bivariate radio-luminosity function for the
early type galaxies in the complex, and compare it to that of galaxies
in more relaxed environments and in the field;

\noindent
{\it (b)} search for extended radio emission, in the form of {\it relics} 
or {\it radio halos}, associated with the clusters rather than with
individual galaxies and presumably consequences of merging precesses;

\noindent
{\it (c)} study the physical properties of the extended radio galaxies
in the complex, in order to derive information on their age, on the
properties of the external medium, and on the projection effects
in the groups or clusters where these sources are located.

The observational data available in the optical band and at X--ray energies
ensure a global analysis of the environment and of the dynamical state of 
the structure.

\subsection{Observations and Data Reduction}

Our starting catalogue is the sample observed at 22 cm. 
The ATCA observations were carried out with a 128 MHz bandwidth, 
and in order to
reduce bandwidth smearing effects, we took advantage of the spectral-line
mode correlation, using 32 channels. The data were reduced using the
MIRIAD package (Sault, Teuben \& Wright 1995), and the analysis of 
the images was carried out with the AIPS package.
The resolution of our survey is 
$\sim 10^{\prime\prime} \times 5^{\prime\prime}$, and the noise ranges from
$70 \mu$Jy to 0.2 mJy/beam. We considered reliable detections all sources
with flux density $S \ge 5 \sigma$, i.e. $S \ge 1$ mJy. 

In Fig. 2 the pointing centres are superimposed to
the optical isodensity contours. The diameter of the circles 
in the figure corresponds $\sim 33$ arcmin, which is the 
size of the primary beam of the ATCA at 22 cm.
A region of $\sim 1$ deg$^2$ around the centre of 
A3556 was mosaiced in 1994-1995 with ten pointing centres (Venturi et al. 
1997), numbered 1 to 10 in Fig. 2. Three more 
pointings were placed south of A3558, at the centre of the group SC 1329
and east of A3562 (N1, N2 and N3 respectively).

Furthemore, in order to completely cover the A3558 complex,
archive ATCA data at the same resolution and band were reduced. These
are the three fields  centered on 
A3558, SC1327 and A3562 (HY31, HY3 and HY22 respectively, from Reid \&
Hunstead). 
Even if not part of the optical region, we included also archive   
observations of A3560 (HY13 in Fig. 2), located half degree
south of the A3558 complex. The proximity of this cluster to the complex
suggests that A3560 is probably infalling toward this structure and possibly 
at the early stages of interaction.
We will not include the analysis of the A3560 image in the present 
paper, and the sources found in this field are not included in the
discussion, however in this paper we will present the peculiar extended 
radio galaxy 
associated with the dominant giant multiple nuclei galaxy in the next section.

As it is clear from Fig. 2 we completely and nearly
homogenously covered the A3558 complex with
our radio survey, going from the western extreme of A3556 to the 
eastern extrem of A3562. The total area covered is 2.9 deg$^2$,
corresponding  to $9 \times 10^{-4}$ sr. 
Such good coverage allows us to study
all the possible environments along the structutre: from the regions 
directly involved  in the merging, where shocks, heating of gas and 
galaxy-galaxy
interactions are expected to play an important role, to the more external
ones, only mildly perturbed by tidal forces.

\subsection{Radio Source Counts and Statistics}

We detected 323 radio sources with $S_{22 cm} \ge 1$ mJy.
Because of the primary beam attenuation, which increases the noise
in the external part of the field, our survey is not complete
down to this flux density limit. Taking into account this effect, 
it is possible to consider our survey complete down to a flux density limit of
2.3 mJy.

Among the 323 radio sources, 69 have an optical counterpart, which corresponds
to 21\%.
27 out of the 69 identifications are in the redshift range of the Shapley 
Concentration, one
is a foreground spiral galaxy (v = 9066 km sec$^{-1}$) and two are 
background objects with a recession velocity v $\sim 58000$ km sec$^{-1}$.
Most of the 16 optical identifications brighter than M$_B$ = 18.5 and
without redshift information  
are very likely to be part of the supercluster, so the number of 
radio galaxies
in the A3558 complex may increase once more redshifts are available in
this region.  The situation is summarised in Fig. 3.

\begin{figure}
\centerline{\vbox{
\psfig{figure=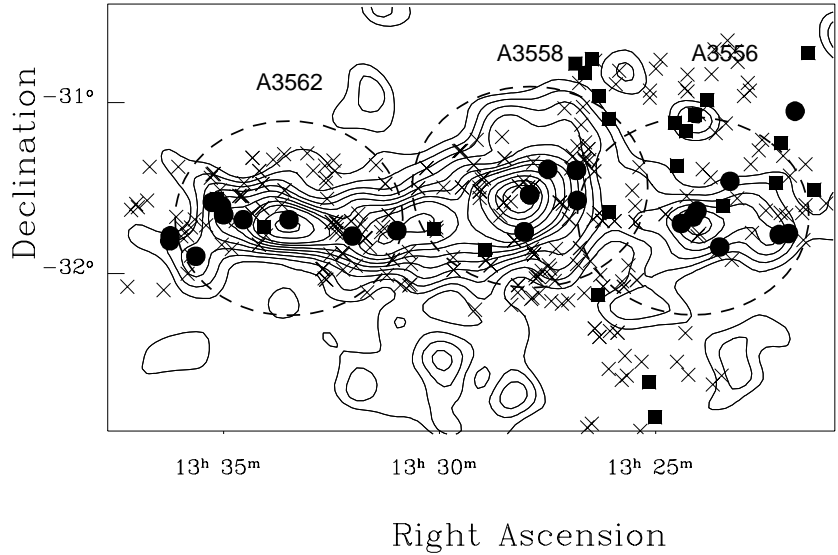,width=0.9\hsize}
}}
\caption[]{The radio sources detected are superimposed to the optical
isodensity contours of the A3558 complex. Filled circles represent radio 
galaxies with known
redshift, filled squares represent radio galaxies without redshift
information and crosses represent radio sources without optical counterpart.}
\end{figure}

\noindent
We can make very preliminary considerations.
A first glance at Fig. 3 reveals that the radio sources are not 
uniformly distributed in the A3558 complex. 
Comparison between Fig. 2 and Fig. 3 ensures us that gaps in the
distribution of the radio sources do not correspond to uncovered
regions in our survey.  
Moreover, 
from Fig. 3 it is also clear that the radio galaxies in the complex
are not clustered in the same way as the galaxy density peaks.
We point out that the 
inner 20 arcmin in A3558, the richest and most massive cluster in the 
chain, contain
only two faint radio galaxies, while the much poorer cluster A3556 and the
outermost region of A3562 contain a large number of radio galaxies.
The peculiarity of A3556 at radio wavelengths was presented and discussed
in Venturi et al. 1997.

\subsection{Extended Radio Galaxies}

Four of the 27 radio galaxies belonging to the A3558 complex
exhibit extended radio emission. In addition, very extended and peculiar
emission is associated with the dominant galaxy in the A3560 cluster.
Their global properties are reported in Table 1.

\begin{table}
\caption{Properties of the Extended Galaxies in the A3558 Complex} \label{tbl-1}
\begin{center}\scriptsize
\begin{tabular}{lllllcllc}
Source & RA$_{J2000}$ & DEC$_{J2000}$ & LogP$_{1.4 GHz}$ & Radio & Optical & 
Velocity & b$_j$ & Cluster \\ 
Name & & & W Hz$^{-1}$ & Type & ID & km s$^{-1}$ & & \\
\tableline
J1322$-$3146 & $13~22~06~~$ & $-31~36~18$ & 22.15 & WAT & E  & 14254 & 14.7 &
A3556 \\ 
J1324$-$3138 & $13~24~01~~$ & $-31~38~~~$ & 23.05 & NAT & E  & 15142 & 15.6  &
A3556 \\
J1332$-$3308 & $13~32~25~~$ & $-33~08~~~$ & 24.38 & WAT & cD & 14586 & 15.6  &
A3560 \\
J1333$-$3141 & $13~33~31.7$ & $-31~41~04$ & 23.33 & NAT & E  & 14438 & 17.3 &
A3562 \\
J1335$-$3153 & $13~35~42.6$ & $-31~53~54$ & 22.60 & FRI & E  & 14385 & 16.0 &
A3562 \\
\end{tabular} 
\end{center}
\end{table}

All these extended sources are associated with bright elliptical galaxies, and
their radio power range is typical of low luminosity FRI radio galaxies
(Fanaroff \& Riley, 1974), as it is commonly found in clusters of galaxies.

\subsubsection{The wide-angle tail J1322$-$3146.}

This radio galaxy was presented and commented briefly on in Venturi 
et al. 1997.
We remind here that it is very unusual, in that it 
exhibits a {\it wide-angle tail} morphology, all embedded in the optical
galaxy, despite the large projected distance from the centre of A3556
(26 armcin, corresponding to 1 h$^{-1}$ Mpc). Wide-angle tail radio galaxies 
are always found at the centre of clusters: this large distance from the
cluster
centre raised the question of what bent the tails. 
Reanalysis of ROSAT archive data and of the Rosat All Sky Survey
(Kull \& Boheringer 1998)
shows that the  gas distribution in the A3558 complex follows closely the
distribution of the optical galaxies reported in Fig. 2 and 3, and 
has a low surface brightness extension which reaches 
the location of J1322$-$3146. A detailed study of this
radio galaxy, including its dynamics and comparison with the properties
of the intergalactic medium as derived from the X--ray data, will be
presented elsewhere.

\subsubsection{The relict radio galaxy J1324$-$3138.}

A multiwavelength detailed study of this source was presented in Venturi
et al. 1998a. Here we summarise the most important conclusions, and
present further considerations on the geometry of the A3556 core. 
 
J1324$-$3138 (see Figs. in Venturi et al. 1998a) is characterised by a steep 
spectrum both in the extended emission ($\alpha = 1.3$ in the range  843 MHz - 
4.9 GHz) and in the ``nuclear'' component coincident
with the optical counterpart. The source has low surface brightness and
lacks polarised emission at any frequency. Its internal pressure
and magnetic field are lower than in typical radio galaxies in the same power
range, and are intermediate between what is found along the tails of
radio sources in galaxies in clusters, and in the few known relic
source. By fitting the spectra of the extended and nuclear components
with a model taking into account reisotropisation of electrons 
(Jaffe \& Perola 1973),  we found that the age of last acceleration
of the emitting electrons is $\sim 10^8$ yrs.
We drew the conclusion that this radio galaxy is a
{\it dead} tailed source, in which the nuclear activity
has switched off. The final evolution of the source is presently
dominted by synchotron losses. 
We interpreted the lack of polarisation assuming 
that the source is seen through a dense screen of gas, 
and concluded that
it is located beyond the core of A3556 and that a major
merging event between the core of A3556 and the subgroup hosting this
radio galaxy has already taken place. 
Our suggestion is that merging
triggered the radio emission in the associated optical galaxy. The old age 
derived for this source is consistent with the timescale of merging.
By estimating the distance of the shock front of the merging from the centre 
of A3556 (Kang et al. 1997, Ensslin et al. 1998), and
comparing with the projected one, we found that the viewing angle is
$\sim 5^{circ}$ implying a maximum dimension for the source
of $\sim$ 1 h$^{-1}$ Mpc.

\subsubsection{The complex radio galaxy J1332$-$3308.}

Reduction of the archive 22 cm ATCA data centered on A3560 (field HY13 in 
Fig. 3),
at the resolution of $10.7^{\prime\prime} \times 6.0^{\prime\prime}$
revealed the presence of a S$_{22 cm}$ = 943 mJy source,
misplaced from the centre of the multiple nuclei cD galaxy located at the 
cluster centre and characterised by a very unusual morphology. 

In order to understand the morphology of this source, and to 
study how it relates 
to the dominant cluster galaxy, we asked for a VLA ad-hoc 
short observation at 20 cm and 6 cm, with the array in the
hybrid configuration BnA,
suitable for the low declination of the source. The results of these
observations are given in Figs. 4a and 4b, where the VLA images
at 20 cm and 6 cm respectively are superimposed on the DSS optical image.

\begin{figure}
\centerline{\vbox{
\psfig{figure=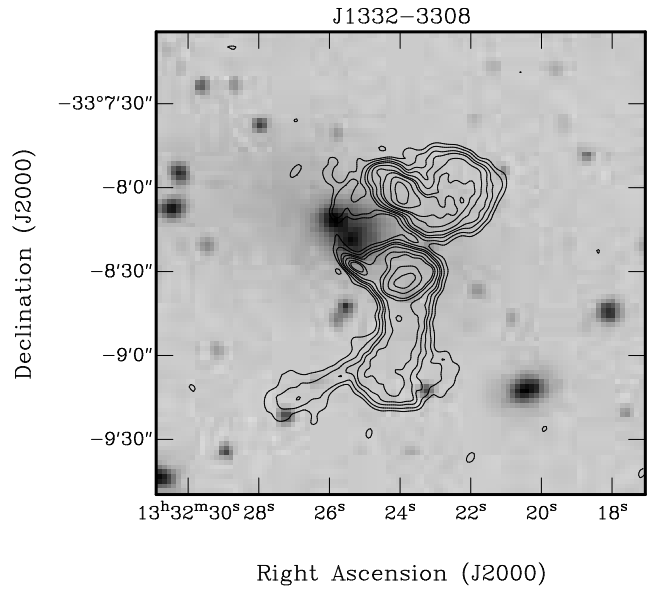,width=0.7\hsize}
\psfig{figure=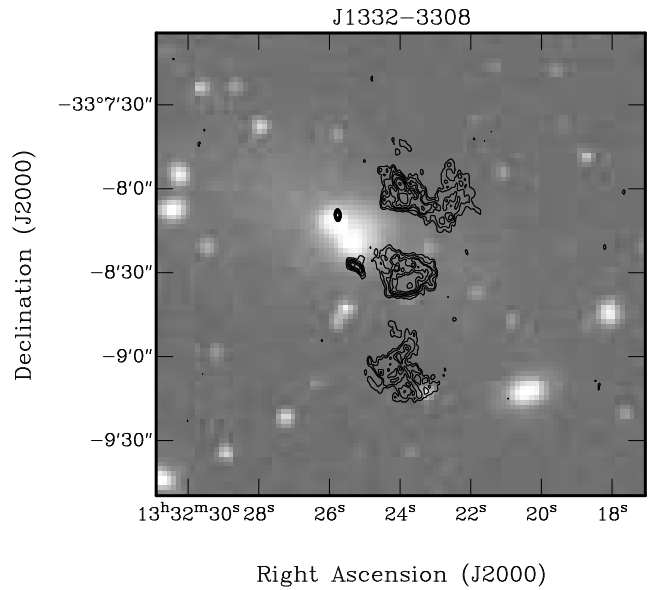,width=0.7\hsize}
}}
\caption[]{(a) Upper. 20 cm VLA full resolution image of J1332$-$3308, 
superimposed on the DSS optical image. The resolution is 
$4.2^{\prime\prime} \times 3.7^{\prime\prime}$, in p.a. 61$^{\circ}$.
(b) Lower. 6 cm VLA image of J1332$-$3308 obtained with natural weight. 
The resolution is 
$2.1^{\prime\prime} \times 1.2^{\prime\prime}$, in p.a. 5.7$^{\circ}$.}
\end{figure}

As it is clear from Figs. 4a and 4b, the morphology of J1332$-$3308 is
complex, and it is suggestive of two distinct components.
The 20 cm emission departs from the northernmost optical nucleus,
in the shape of a wide-angle tail radio galaxy. The nucleus of the
emission is visible in the 6 cm image, coincident
with the norhernmost brightest optical nucleus and this reinforces the idea 
that this radio galaxy is a wide-angle
tail source associated with an active nucleus, despite the lack of a
visible jet in the high resolution map.

The ridge of emission south of the secondary optical nucleus, well
visible at 6 cm, and the southernmost amorphous extended emission
are difficult to be interpreted. It seems unlikely that they are related 
to the wide-angle tail component. In the 6 cm map the ridge seems to
point to the southernmost diffuse component. 
The overall structure of J1332$-$3308 is reminescent of 3C338, seen under
a different viewing angle. 3C338 is an extended classical double source
associated with the multiple nuclei galaxy NGC6166, located at the centre
of the Abell cluster A2199. The steep spectrum ridge visible in 3C338
south of the location of the presently active nucleus was interpreted
by Burns et al. 1993 as the remnant of a previous activity in the
galaxy, associated with a different optical nucleus. 
The resolution and u-v coverage of our observations at 20 cm and 6 cm are
too different to carry out a spectral index
study, which is essential to derive the intrinsic properties of
the various components, including their age, and to test 
our preliminary hypothesis that J1332$-$3308 is a {\it reborn} radio
galaxy, similarly to 3C338. We are currently engaged in
a detailed multifrequency study of this source, in order to disentangle 
its nature.

\subsubsection{The head-tail source J1333$-$3141 and the double J1335$-$3135.}

22 cm maps of these two radio galaxies were presented in Venturi et al. 1998b.
They both belong to A3562, the easternmost cluster in the A3558 complex.

The head-tail source J1333$-$3141 is associated with a 17th magnitude 
elliptical galaxy
located at $\sim 1^{\prime}$ from the centre of A3562, and  it is possibly
orbiting around the cD galaxy in the potential well of the cluster.
The total extent
of the source is $\sim 1^{\prime}$ corresponding to $\sim$ 40 kpc.
The tail is straight up to $\sim 20^{\prime}$ from the
nucleus, then the bending becomes relevant and the emission diffuse, 
suggesting that
the two jets forming it (not visible in our map because of resolution
effects) open at this distance from the core. 
In order to derive some global properties of this radio galaxy
we studied in the wavelength range 13 - 22 cm.
Fig. 5 shows the 13 cm flux density contours of this
source, superimposed on the DSS optical image.

\begin{figure}
\centerline{\vbox{
\psfig{figure=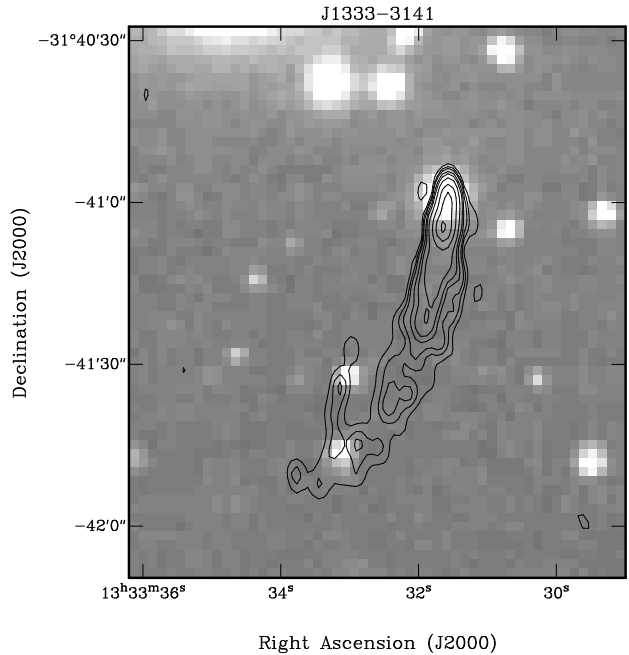,width=0.7\hsize}
}}
\caption[]{13 cm ATCA image of J1333$-$3141, superimposed
on the DSS optical image. The resolution is 
$5.4^{\prime\prime} \times 3.2^{\prime\prime}$, in p.a. 0.7$^{\circ}$.}
\end{figure}
\begin{figure}
\centerline{\vbox{
\psfig{figure=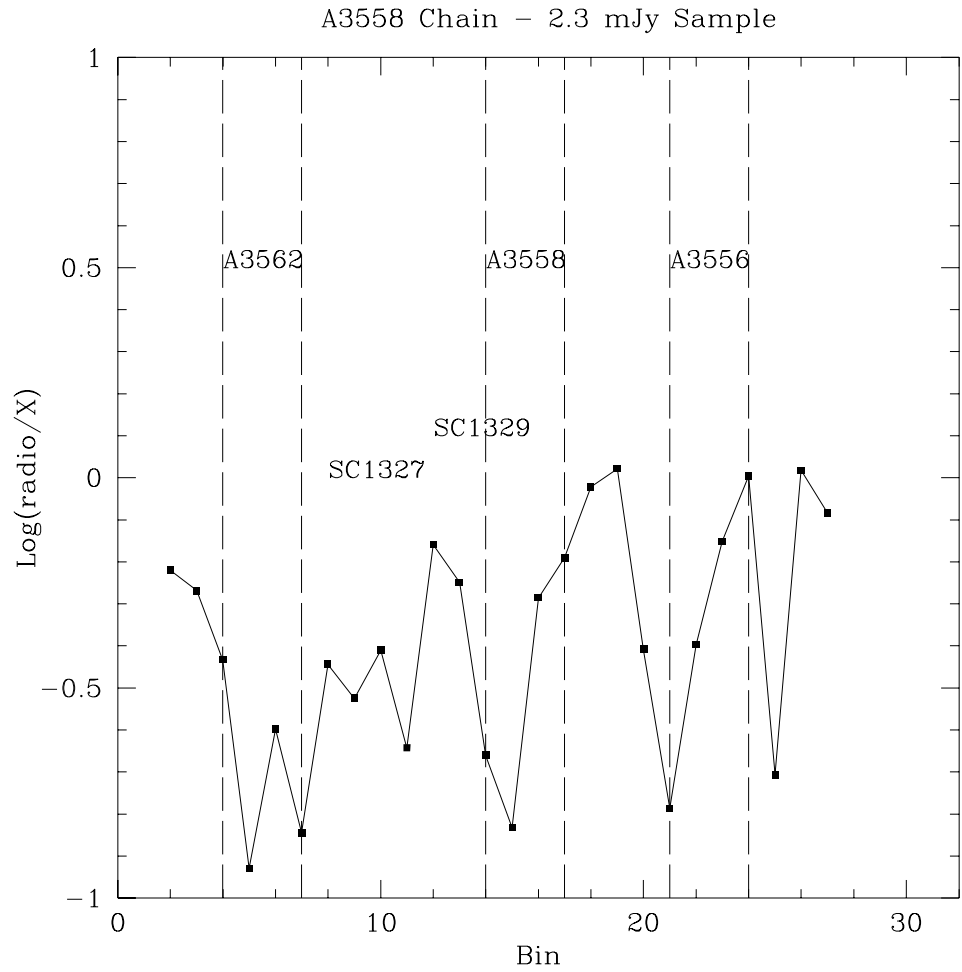,width=\hsize}
}}
\caption[]{Ratio between the radio source counts and X--ray counts
in the A3558 complex for the complete radio sample . 
The bins are the same as in Kull \& Boehringer 1998.}
\end{figure}

\noindent
The total spectral index $\alpha_{13 cm}^{22 cm}$ in this source
at the resolution our the 22 cm image survey is 0.9. This should probably
be considered an upper limit since the 13 cm u-v plane is not well
covered at the short spacings, and extended flux from the tail might
have been missed in our image. With this preliminary value of 
the spectral index we derived the equipartition magnetic field and
internal pressure in the source, computed assuming a filling factor
$\phi$=1, k=1 (ratio between protons and electrons) and integrating
over a frequency range 10 MHz - 100 GHz. We obtained H$_{eq} = 3 \mu$G
and P$_{eq} = 5.5 \times 10^{-13}$ dyne cm$^{-2}$. As stated above,
these values should probably be considered lower limits, however, though
preliminary, they are consistent with the average intrinsic physical quantities
of the extended emission in cluster galaxies, and are significantly different
from what is derived for J1324$-$3138, the extended radio galaxy at the
other end of the A3558 complex. 

J1335$-$3153 is a double source located at $\sim 12^{\prime}$ from the
centre of A3562, in the east direction. The radio morphology is asymmetric,
with the western lobe longer and more distorted than the eastern one.
It is possible that this source has a bent or distorted morphology
and that we see it face-on.
A more detailed study and comparison with the properties of the
intergalactic medium as derived from the X--ray data is in progress.

\section{Radio source distribution and X--ray emission}

It is now well established that the gas distribution in the A3558
complex perfectly matches the distribution of the galaxy density,
with galaxy density peaks coincident with gas density and temperature peaks.
This result 
further reinforce the physical connection of gas and galaxies in the
Shapley Concentration core (Bardelli et al. 1994, Bardelli et al. 1996,
Ettori et al. 1997, Kull \& Boehringer 1998).
Since the 22 cm radio survey presented here covers the whole extent of
the optical and X--ray data, we carried out a comparison between 
the X--ray distribution and the radio source distribution, in order
to see if the radio galaxies also peak in galaxy and gas density
peaks.

We referred to Kull \& Boehringer 1998 for our comparison, and 
binned the radio sources in our survey in the same 30 bins chosen 
by those authors to derive the gas density profile along the A3558 complex.
We computed and plotted the ratio between the radio source counts and 
the X--ray luminosity (in counts sec$^{-1}$ arcmin$^{-2}$) in each bin,
both for the uncomplete (but deeper) 1 mJy sample and for the complete
2.3 mJy sample. We comment here that the sources in our survey 
obviously include the radio background. However, since it can be considered
constant over the whole region of the A3558 complex it
does not affect the shape of the radio source distribution.
Our preliminary result, shown in Fig. 6, indicates that there is a deficiency 
of radio sources in the the X--ray peaks, i.e. in coincidence with the
core of A3558 and A3562. Alternatively our plot can be interpreted as 
an excess of radio sources in those regions where ongoing merging is
thought to be present.

\section{Preliminary Considerations}

The most strinking result of our 22 cm survey is the non-uniform
distribution of radio sources in the A3558 complex. The most
massive and relaxed cluster in the chain, A3558, exhibits very little
radio emission, while the two extremes of the chain, where interactions
and merging are supposed to dominate the dynamics, are populated by
a large number of radio galaxies. A peak in the radio source distribution is
also found in coincidence with the two groups SC1327 and SC1329, known to 
be dynamically active.
The apparent anticorrelation between the radio source counts and the
X--ray emission in the chain also points towards the same considerations.
Our radio counts versus X--ray emission preliminary analysis on the 
A3558 complex
suggests that the radio galaxies seem to avoid either very high gas density
environments or evolved clusters. At this stage we cannot discriminate 
between these two possibilities. Comparison and analysis of literature
data is essential to further study this effect.

The study of the extended radio galaxies in this complex has revealed
the presence of a complex source associated with the multiple nuclei
dominant galaxy in A3560, which could be the result of a restarted 
radio activity, possibly connected with merging processes.

\end{document}